\documentclass[aps,twocolumn,showpacs]{revtex4}
\makeatletter
\newcommand{\rmnum}[1]{\romannumeral #1}
\newcommand{\Rmnum}[1]{\expandafter\@slowromancap\romannumeral #1@}
\makeatother
\usepackage{color}
\usepackage{graphicx}
\usepackage{dcolumn}
\usepackage{bm}
\usepackage{CJK}
\usepackage{amsfonts}
\usepackage{psfrag}
\usepackage{wrapfig}
\usepackage{subfigure}
\usepackage{makeidx}
\usepackage{multirow}
\usepackage{epsf}
\usepackage{hyperref}
\usepackage{amsmath}
\usepackage{cases}

\begin{document}

\title{Characteristics of optical multi-peak solitons induced by higher-order effects in an erbium-doped fiber system}
\author{Yang Ren$^{1,2}$}
\author{Zhan-Ying Yang$^{1,2}$}\email{zyyang@nwu.edu.cn}
\author{Chong Liu$^{1,2}$}\email{nwudavid@163.com}
\author{Wen-Hao Xu$^{1,2}$}
\author{Wen-Li Yang$^{2,3}$}
\address{$^1$School of Physics, Northwest University, Xi'an 710069, China}
\address{$^2$Shaanxi Key Laboratory for Theoretical Physics Frontiers, Xi'an 710069, China}
\address{$^3$Institute of Modern Physics, Northwest University, Xi'an 710069, China}

\begin{abstract}
We study multi-peak solitons \textit{on a plane-wave background} in an erbium-doped fiber system with some higher-order effects,
which is governed by a coupled Hirota and Maxwel-Bloch (H-MB) model. The important characteristics of multi-peak solitons induced by the higher-order effects,
such as the velocity changes, localization or periodicity attenuation, and state transitions, are revealed in detail.
In particular, our results demonstrate explicitly that a multi-peak soliton can be converted to an anti-dark soliton when
the periodicity vanishes; on the other hand, a multi-peak soliton is transformed to a periodic wave when the localization
vanishes. Numerical simulations are performed to confirm the propagation stability of multi-peak solitons riding on a plane-wave background.
Finally, we compare and discuss the similarity and difference of multi-peak solitons in special degenerate cases of the H-MB system with general existence conditions.
\end{abstract}
\pacs{05.45.Yv, 42.65.Tg, 02.30.Ik}
\maketitle

\section{INTRODUCTION}
Nonlinear waves propagating in an erbium-doped fiber system have attracted special attention,
since the resonant absorption of the erbium-doped two levels system is a good solution to the optical losses \cite{n,n1,n2,n3,h1,e1,e2,e3,e4}.
One of the most well-known examples is the amplification and reshaping of optical soltion (bright soliton on a zero background) in a long-haul optical
communication through fibers with an active region doped with erbium atoms.
In this case, nonlinear wave propagation could possess both the effects due to a mix of silica and erbium impurities.
More specifically, the silica material gives the group velocity dispersion and self-phase modulation effects, which is governed by the nonlinear Schr\"{o}dinger (NLS) equation; whereas Er impurities contribute to the self-induced transparency (SIT), which is described by the Maxwell-Bloch (MB) model \cite{n,n1,n2,n3,h1,e1,e2,e3,e4}.
As a result, the constraint to the NLS soliton namely the optical losses can be somewhat compensated
with the effect of SIT. In general, the dynamics of these waves in the erbium-doped fibers is described
by the coupled system of the NLS and the MB equations \cite{n,n1,n2,n3}.
It should be emphasized that recent experiments have confirmed guided wave
SIT soliton formation and propagation by employing a few meters of erbium-doped fiber \cite{e1,e2,e3,e4}.
Moreover, the coexistence of the NLS soliton and SIT soliton has already been realized experimentally \cite{e3,e4}.

In addition to the classical solitons on a zero background,
nonlinear waves on a plane-wave background have
recently been the subject of intensive investigations in nonlinear optics \cite{r1,r2}. Generally, nonlinear waves riding on a plane-wave background
exhibit the onset of modulation instability, which are known as rogue waves and breathers \cite{r1,r2}. In contrast to the soliton with stable structural feature, breathers and rogue waves are localized structures with unstable characteristic. Breathers are the localized breathing waves with a periodic profile in a certain direction; rogue waves are rare, short-lived, and localized in both space and time, which are some special cases of breathers \cite{r1,r2}.
The corresponding analogs of breathers and rogue waves in the coupled NLS-MB system have been revealed \cite{b1,b2,b3,b4,b5,b6,b7}.
However, it was found recently that nonlinear waves on a plane-wave background in the coupled NLS-MB system can
display structural diversity \cite{b6}.
Remarkably, an interesting type of solitons on a plane-wave background with a striking multi-peak structure
whose feature is completely different from rogue waves and breathers is revealed \cite{b6}. It shows that the multi-peak soliton exhibits both localization and periodicity along the transverse distribution arising from the interaction between NLS and MB components, which cannot exist in the standard NLS system.
One should note that the NLS-MB model is available for the pulse propagation in the picosecond regime \cite{n,n1,n2,n3}.

But for ultrashort pulses (duration shorter than 100fs), higher-order effects come into play and impact the characters of localized waves significantly \cite{t1,t2,t3,t4}.
The resulting higher-order NLS models thus describe the dynamics of nonlinear waves with higher accuracy than the standard NLS equation.
Recent studies have demonstrated that the standard solitons do exist in a series of higher-order NLS hierarchies, and the
higher-order effects impact the velocity of solitons \cite{s1,s2}.
On the other hand, the existence of the multi-peak solitons in an erbium-doped fiber system with some higher-order effects is demonstrated \cite{b7}.
However, important characteristics of multi-peak solitons induced by the higher-order effects,
such as the velocity changes, localization or periodicity attenuation, and state transitions, have not been revealed so far, to our knowledge.

In this paper, we study the characteristics of multi-peak solitons induced by higher-order effects in an erbium-doped fiber system,
which is governed by the coupled Hirota and Maxwel-Bloch (H-MB) model.
Our results show that the higher-order effects influence not only the velocity but also the
localization and periodicity of the multi-peak soliton, which can be identified by the relation between the background frequency $q$ and the coefficient of higher-order terms $\beta$. The propagation stability of multi-peak solitons riding on a plane-wave background is confirmed numerically.

\section{The H-MB system and exact multi-peak soliton solution}
We consider ultrashort pulses propagating a resonant erbium-doped fiber system with the
important higher-order effects above governed by a coupled system of the H-MB model \cite{h1}
\begin{eqnarray}
\label{equ:1}
    &&E_z=i\alpha(E_{tt}+2|E|^2E)+\beta (E_{ttt}+6|E|^2E_t)+2P, \nonumber\\
    &&P_t=2i\omega P+E \eta ,\nonumber\\
    &&\eta_t=-(EP^*+PE^*) ,
\end{eqnarray}
where $E(z,t)$ is the normalized slowly varying amplitude of the complex field envelope; $P(z,t)$ is the measure of
the polarization of the resonant medium, which is defined by $P = v_1v_2^*$; $\eta(z,t)$ denotes the extent of the population
inversion, which is given by $\eta=|v_1|^2-|v_2|^2$, $v_1$ and $v_2$ are the wave functions of the
two energy levels of the resonant atoms; $\alpha$ relates to the group velocity dispersion and the self-phase modulation;
$\beta$ (a small value) which is responsible for the higher-order terms including the third-order dispersion, self-steepening,
and nonlinear response effects; $\omega$ is the carrier frequency, and the index $*$ denotes complex conjugate.

In order to study multi-peak solitons on a plane-wave background in the H-MB model, we first introduce the following
background solution with a generalized form
\begin{eqnarray}
\label{equ:2}
    E_1=ae^{i \theta },~
    P_1=i k E_1,~
    \eta_1=\omega k -q k/2,
\end{eqnarray}
where $\theta=qt+\nu z$, $\nu=2k-q^2(\alpha+\beta q)+2a^2(\alpha+3\beta q)$, $a$ and $q$ represent the amplitude and
frequency of background electric field, respectively, and $k$ is a real parameter which is related to the background amplitude of the $P$ component.

If the background amplitudes vanish, i.e., $a=0$, $k=0$, Eq.\ref{equ:2} reduces to the trivial solution, which can be used
to generate standard bright soliton solutions \cite{h1}. Correspondingly, the impact of the higher-order effects on the standard bright
solitons and soliton interactions has been studied \cite{h1,b4}. It is demonstrated that the higher-order effects do not affect the soliton
structure and they affect merely the velocity of the waves.

Here in order to study multi-peak solitons on the plane-wave background
(\ref{equ:2}) in the H-MB model, we shall keep $a,k\neq0$.
On the other hand, we pay our attention to multi-peak structures in electric field, i.e., the pulse propagation in the $E$ component.
We omit the results in the $P$, $\eta$ components, since their types of nonlinear waves are similar to the ones in the $E$ component
with the same initial physical parameters.

By using the Darboux transformation technique \cite{m},
the general first-order exact multi-peak soliton solution on the plane-wave background in the $E$ component is given
\begin{eqnarray}
\label{equ:3}
    E=E_1 \left\{1+\frac{8b_1m_1[\sin{(\sigma+\psi_1)}+i\sinh{(\tau-i\psi_1)}]}
    {m_3\sinh(\tau-i\psi_3)-m_2\sin{(\sigma+\psi_2)}}\right\}.
\end{eqnarray}
It is evident that the solution is formed by a nonlinear superposition of the hyperbolic function $\sinh{\tau}$ and
trigonometric function $\sin{\sigma}$ on the background $E_1$. This unique nonlinear superposition signal exhibits
the characteristics of the nonlinear structures on the nonvanishing background, which is expressed as:
\begin{eqnarray}
\tau&=&\zeta(t+V_1z),~~~~~~~~\sigma=\gamma(t+V_2z), \nonumber \\
V_1&=&v_1+\gamma b_1 v_2/\zeta,~~~~V_2=v_1-\zeta b_1 v_2/\gamma, \nonumber \\
v_1&=&\alpha(2a_1-q)+\beta(4b_1^2+2a^2-q^2-4a_1^2+2aq) \nonumber \\
&+&k(\omega+a_1)/[b_1^2+(a_1+\omega)^2], \nonumber \\
\label{equ:c}
v_2&=&2(\alpha+\beta q-4a_1 \beta)-k/[b_1^2+(a_1+\omega)^2], \\
\zeta&=&\left(\sqrt{\chi_1^2+\chi_2^2}+\chi_1\right)^{1/2}/\sqrt{2}, \nonumber \\
\gamma&=&\pm\left(\sqrt{\chi_1^2+\chi_2^2}-\chi_1\right)^{1/2}/\sqrt{2},\nonumber \\
\chi_1&=&4b_1^2-4a^2-(2a_1+q)^2, \chi_2=-4b_1(2a_1+q),\nonumber
\end{eqnarray}
with the corresponding amplitude and phase notations:
\begin{eqnarray}
m_1&=&\left[(2iq-4b_1+4ia_1)^2+(2i\zeta+2\gamma)^2\right]^{1/2},\nonumber \\
m_2&=&2\left[(\mu_1-\mu_2)^2-\mu_4^2\right]^{1/2},\nonumber \\
m_3&=&2i\left[(\mu_1+\mu_2)^2-\mu_3^2\right]^{1/2},\nonumber \\
\mu_1&=&4b_1^2+4a^2+(2a_1+q)^2,~~\mu_2=\zeta^2+\gamma^2, \nonumber \\
\mu_3&=&4a_1\gamma+4b_1\zeta+2q\gamma,\nonumber \\
\mu_4&=&i(4b_1\gamma-4a_1\zeta-2\zeta q),\nonumber \\
\tan\psi_1&=&(iq-2b_1+2ia_1)/(i\zeta+\gamma),\nonumber \\
\tan\psi_2&=&(\mu_1-\mu_2)/i\mu_4,~~\tan\psi_3=-(\mu_1+\mu_2)/i\mu_3.\nonumber
\end{eqnarray}
Here, $¡®$$\pm$$¡¯$ in $\gamma$ depends on $\chi_2\leq$0 and $\chi_2>$0, respectively. The above solution depends on the
background wave amplitudes $a$, $k$, the background wave frequency $q$, the real nonzero parameters $a_1$ and $b_1$,
the carrier frequency $\omega$,  and the real parameters $\alpha$ and $\beta$.

One should note that the multi-peak soliton exists under the condition $v_2=0$ [see Eq. (\ref{equ:c})], which result in that the solution features
both hyperbolic and trigonometric with the same velocity $v_1$. Thus the multi-peak soliton exhibits the localization and
periodicity along the transverse distribution $t$ on a plane-wave background [see the following intensity figures].
Additionally, recent studies demonstrate that the multi-peak soliton can exist in many different nonlinear physical systems \cite{a1,a2,w,w1,l}.
Thus it could be regarded as another nonlinear mode on a plane-wave background whose feature is completely different from the well-known breather and rogue wave.

In the following we will study the characteristics of multi-peak solitons in the H-MB system induced by the higher-order
effects based on the exact solution with the existence condition $v_2=0$.
Remarkably, we find that the relation between the background frequency $q$ and the coefficient of higher-order terms $\beta$ plays a critical rule in the impact of higher-order effects on the multi-peak solitons.
It follows that the higher-order effects influence not only the velocity but also the localization and periodicity of the multi-peak soliton.

\section{Characteristics of multi-peak solitons induced by higher-order effects}

\subsection{Multi-peak solitons with different velocities}
\begin{figure}[htb]
\centering
\subfigure{\includegraphics[height=42.5mm,width=42.5mm]{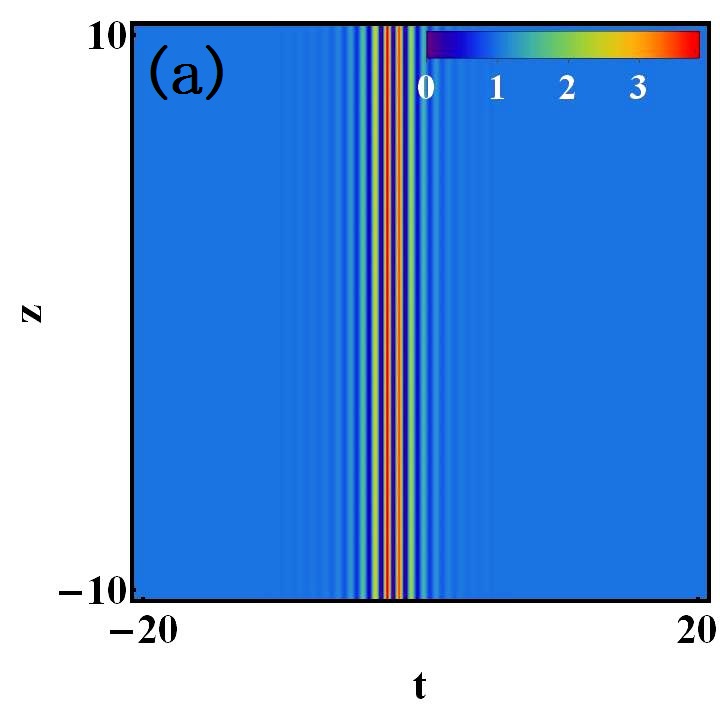}}
\subfigure{\includegraphics[height=42.5mm,width=42.5mm]{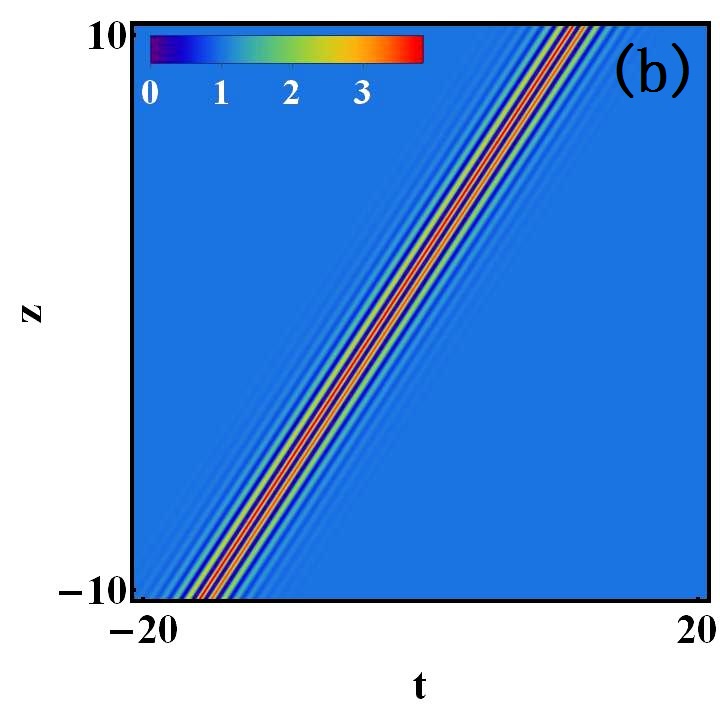}}
\subfigure{\includegraphics[height=42.5mm,width=42.5mm]{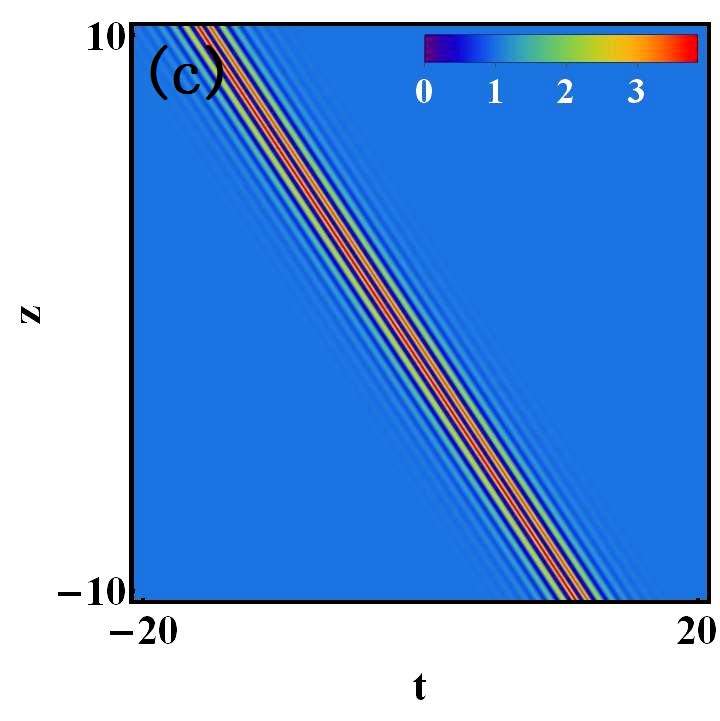}}
\subfigure{\includegraphics[height=42.5mm,width=42.5mm]{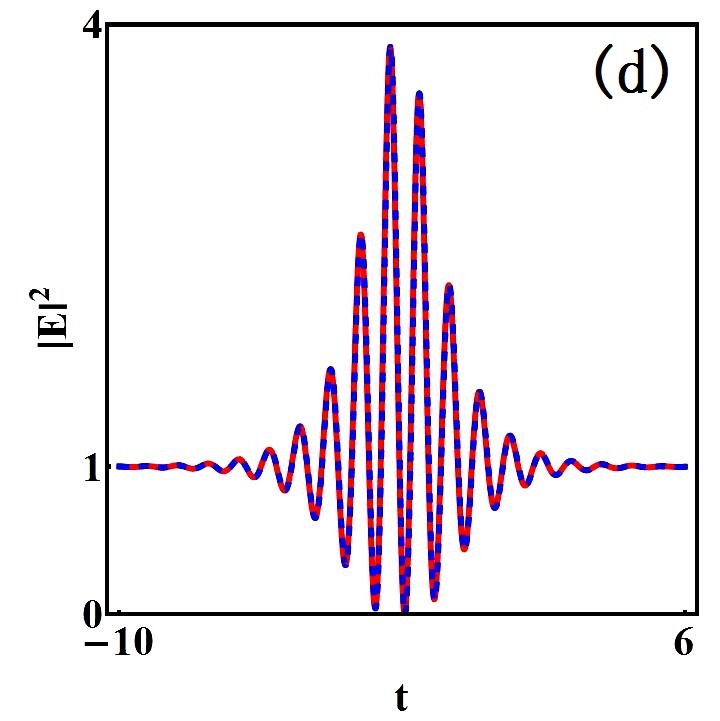}}
\caption{Different velocities of multi-peak solitons with an identical intensity structure induced by higher-order
effects in the H-MB system for the choice of $\beta$,
(a) the corresponding intensity distributions ($I = |E|^2$) with $\beta=0$, (b) $\beta=0.1$, (c) $\beta=-0.1$,
(d) the intensity profiles of (a), (b), and (c) at $z=0$. Other parameters are $\alpha=0.5$, $a=1$, $a_1=1$, $b_1=0.5$,
$\omega=1$, $q=2\omega+4a_1$. It is shown that the higher-order effects only give rise to velocity changes of multi-peak solitons,
and they do not affect intensity distributions.}\label{fig.1}
\end{figure}

We first study the case that the higher-order effects only impact the velocity of  the multi-peak soliton with an identical intensity structure.
To this end, we extract the existence condition of multi-peak solitons from Eq. (\ref{equ:c}) ($v_2=0$) as follows:
\begin{eqnarray}
\label{equ:c1}
k=2n[\alpha+\beta(q-4a_1)],~~n=b_1^2+(a_1+\omega)^2.
\end{eqnarray}
Note that, in this case, the parameter $q$ and the coefficient of higher-order terms $\beta$ are two independent parameters.
As a result, for any given background frequency $q$, one can observe the properties of multi-peak solitons with different values of $\beta$.
To better study the impact of higher-order effects on the multi-peak solitons, we first select a stationary multi-peak soliton ($v_1=0$)
in absence of higher-order effects ($\beta=0$), implying $q=2\omega+4a_1$. We then study the property of this stationary wave in presence
of the higher-order effects ($\beta\neq0$).
The corresponding typical intensity [$I(z,t)=|E(z,t)|^2$] distributions are depicted well in Fig. \ref{fig.1}.

Figure \ref{fig.1}(a) exhibits a stationary multi-peak soliton in absence of higher-order effects ($\beta=0$). In this case, the multi-peak solution reduces to the
solution reported in the NLS-MB model. Figures \ref{fig.1}(b) and \ref{fig.1}(c) depict the multi-peak soliton for $\beta>0$ and $\beta<0$, respectively.
It shows that the multi-peak solution possesses opposite (positive and negative) velocity depending on $\beta>0$ and $<0$.
Moreover, the velocity increases as $|\beta|$ increases. Nevertheless, a comparison of the intensity profiles in Figs. \ref{fig.1}(a)-(c)
implies that, the multi-peak soliton has an identical intensity structure, which is shown in Fig. \ref{fig.1}(d). This indicates that the higher-order
effects affect the soliton's velocity and do not impact the wave structure for a fixed value of background frequency $q$.
One should note that the similar impact of higher-order effects on the velocity of other types of nonlinear waves (the standard solitons, breathers and rogue waves) has also been revealed in various higher-order NLS systems \cite{s1,s2,x1,x2,x3,x4,x5}. It is therefore believed that the velocity change of nonlinear waves induced by higher-order effects is one of the universal properties of nonlinear waves in different nonlinear systems in presence of higher-order effects.

\begin{figure}[htb]
\centering
\subfigure{\includegraphics[height=42.5mm,width=42.5mm]{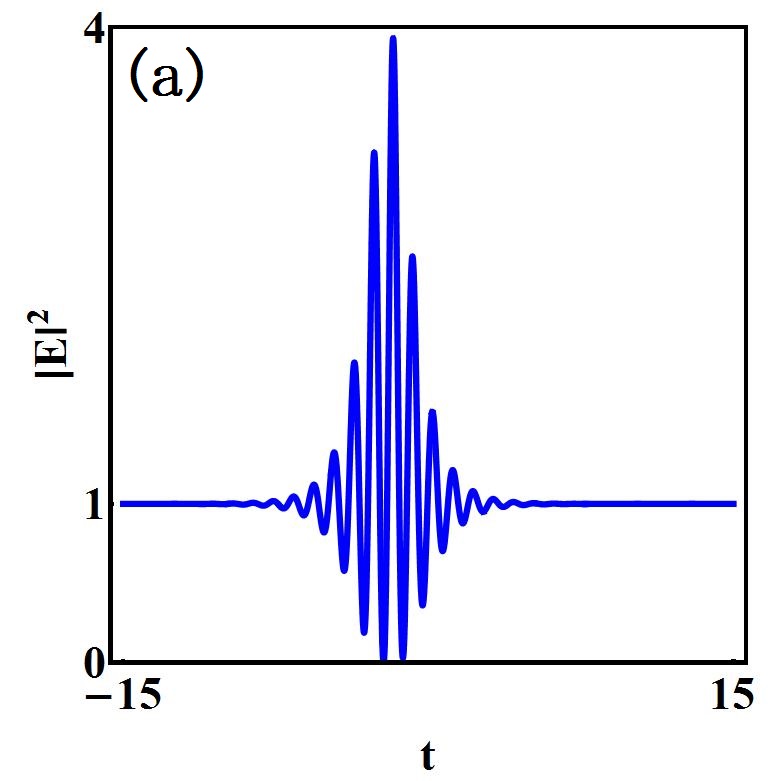}}
\subfigure{\includegraphics[height=42.5mm,width=42.5mm]{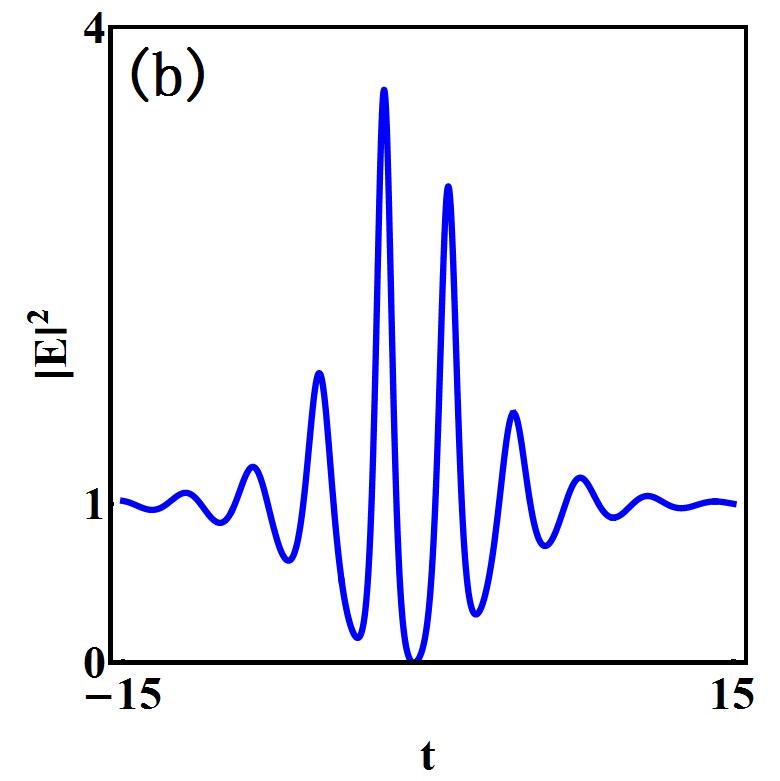}}
\subfigure{\includegraphics[height=42.5mm,width=42.5mm]{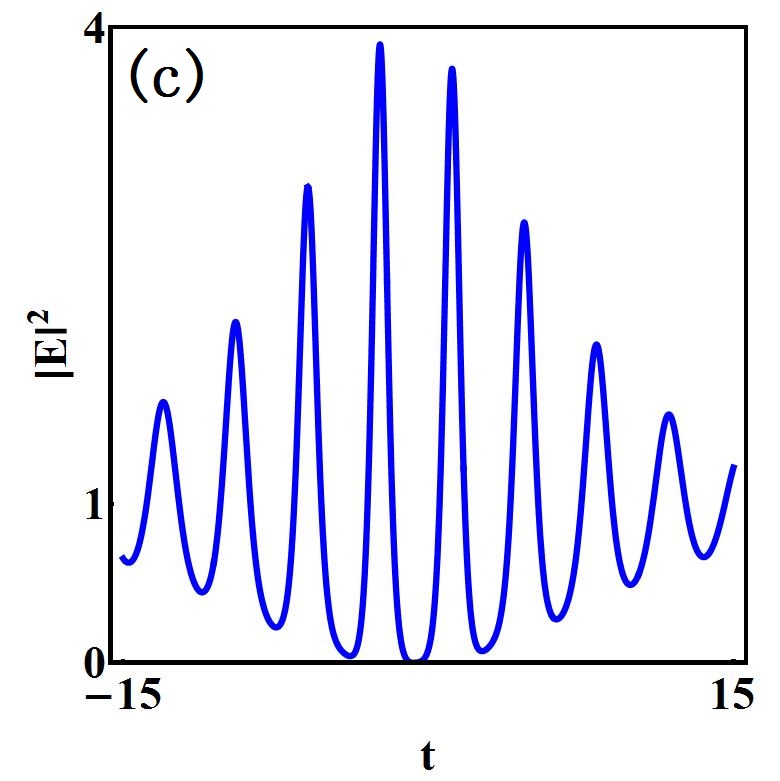}}
\subfigure{\includegraphics[height=42.5mm,width=42.5mm]{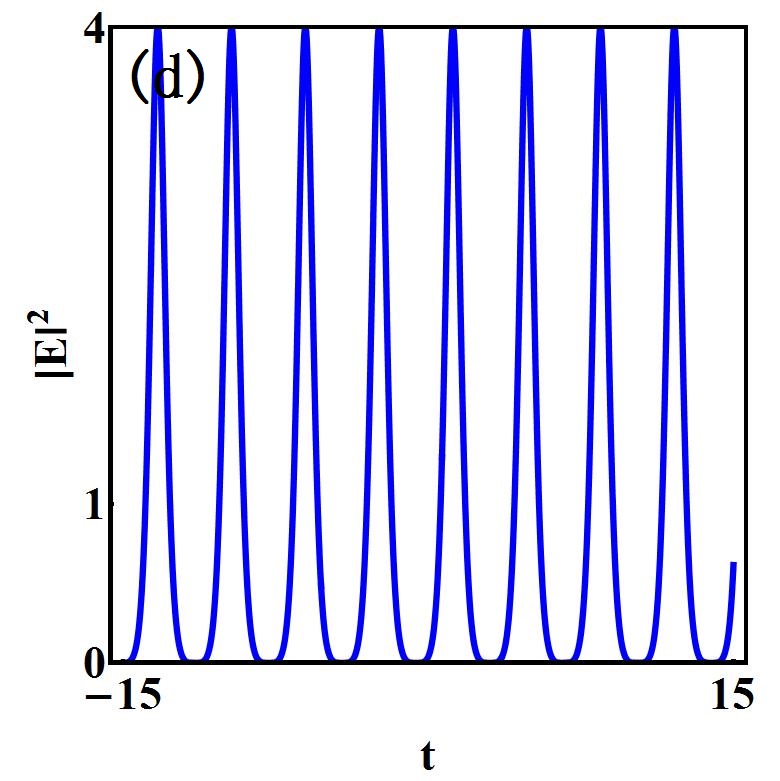}}
\caption{The impact of higher-order effects on the localization of multi-peak solitons in the H-MB system with different values of $\beta$,
(a) the corresponding intensity profiles with $\beta=0.5\beta_s$, (b) $\beta=0.9\beta_s$, (c) $\beta=0.96\beta_s$, (d) $\beta=\beta_s$
with $\beta_s=(2\alpha n-k)/(12a_1n)$. Other parameters are $\alpha=0.5$, $a=1$, $a_1=1$, $b_1=0.5$, $\omega=1$, $k=1$. It is shown that
the localization of multi-peak solitons decreases as $\beta\rightarrow\beta_s$. The localization vanishes when $\beta=\beta_s$, which result
in that the multi-peak soliton is converted to a periodic wave.}\label{fig.2}
\end{figure}

\subsection{Localization or periodicity attenuation and state transitions of multi-peak solitons}
Next, we turn our attention to the case that the higher-order effects influence the localization and periodicity of the multi-peak soliton.
For this purpose, we extract the existence condition of multi-peak solitons from Eq. (\ref{equ:c}) ($v_2=0$) as follows:
\begin{eqnarray}
\label{equ:c2}
&q=4a_1-6a_1\beta_s/\beta,~\beta_s=(2\alpha n-k)/(12a_1n),
\end{eqnarray}
Here the background frequency $q$ is expressed as a function of $\beta$ (thus $\beta_s\neq0$, implying $k\neq2\alpha n$); in particular,
the parameter $\beta$ in this case should be nonvanishing, thus the higher-order effects may play an important role in the localized wave property.

\begin{figure}[htb]
\centering
\subfigure{\includegraphics[height=42.5mm,width=42.5mm]{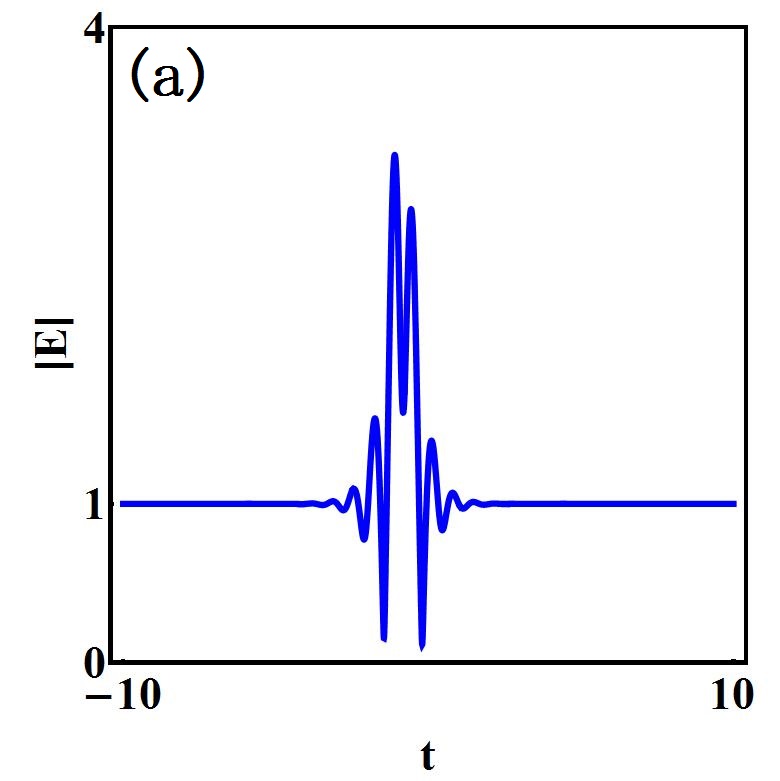}}
\subfigure{\includegraphics[height=42.5mm,width=42.5mm]{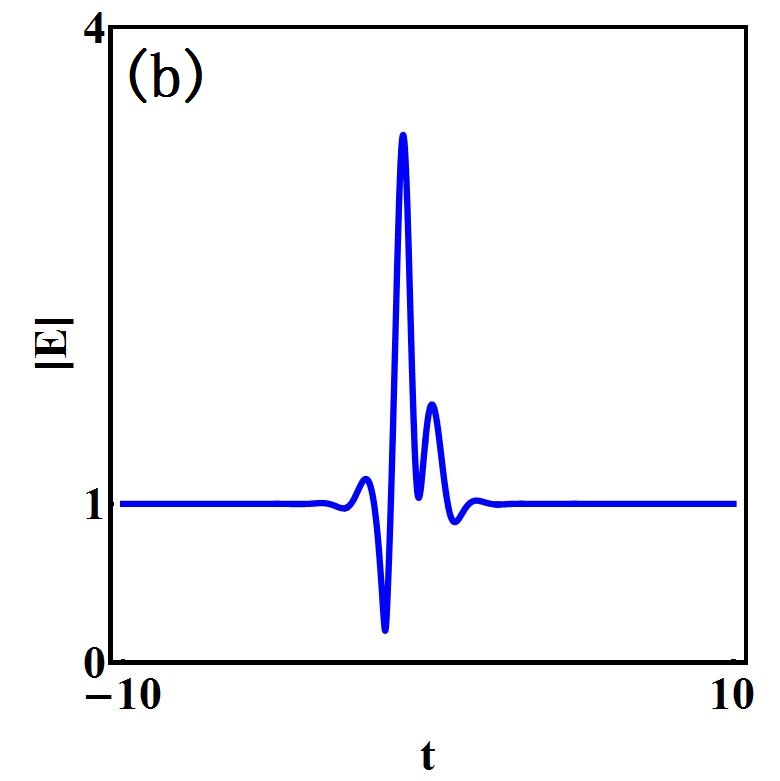}}
\subfigure{\includegraphics[height=42.5mm,width=42.5mm]{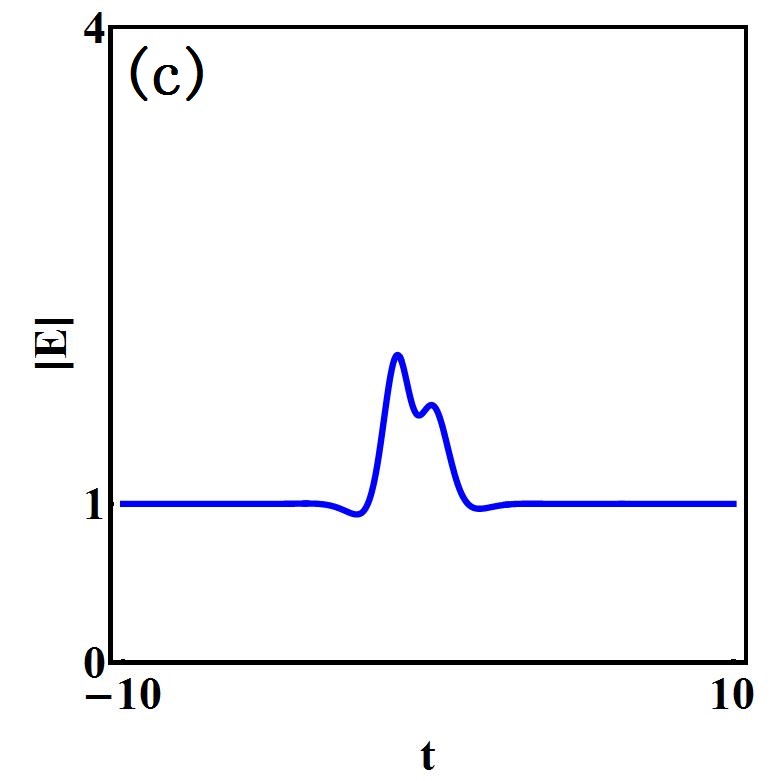}}
\subfigure{\includegraphics[height=42.5mm,width=42.5mm]{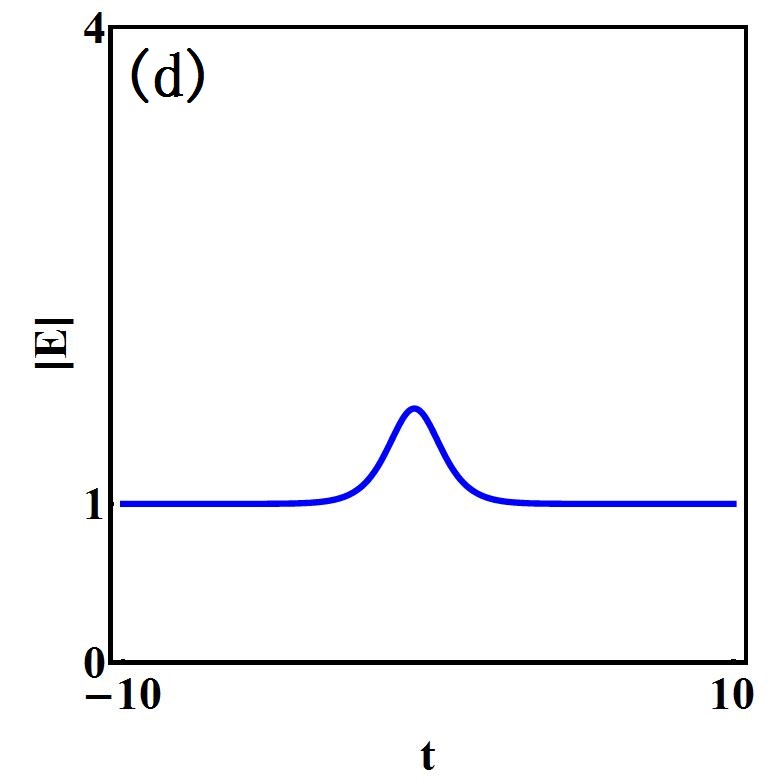}}
\caption{The impact of higher-order effects on the periodicity of multi-peak solitons in the H-MB system with different values of $\beta$,
(a) the corresponding amplitude profiles with $\beta=0.4\beta_s$, (b) $\beta=0.6\beta_s$, (c) $\beta=0.8\beta_s$, (d) $\beta=\beta_s$
with $\beta_s=(2\alpha n-k)/(12a_1n)$. Other parameters are the same as in Fig. 2, but $b_1=1.3$. It is shown that the periodicity of
multi-peak solitons decreases as $\beta\rightarrow\beta_s$. Eventually the periodicity vanishes when $\beta=\beta_s$, which result in that
the multi-peak soliton is converted to an anti-dark soliton.}\label{fig.3}
\end{figure}

Interestingly, we find that the higher-order effects do impact the localization and periodicity of the multi-peak soliton with fine
tuning of $\beta$. Especially, in the particular case of $\beta=\beta_s$, the multi-peak soliton can be converted to a periodic wave
with vanishing localization ($a^2>b_1^2$), and the multi-peak soliton can be transformed to an anti-dark soliton with vanishing periodicity ($a^2<b_1^2$).
In the following, we will study the impact of higher-order effects on the localization and periodicity as $\beta\rightarrow\beta_s$.
The corresponding structure profiles are displayed well in Figs. \ref{fig.2} and \ref{fig.3}.

Figure \ref{fig.2} shows the impact of higher-order effects on the localization of the multi-peak soliton with $a^2>b_1^2$. It is evident to
observe that, as $\beta\rightarrow\beta_s$, the multi-peak soliton has its $t$-direction (transverse) localization gradually decreasing and its periodicity increases.
When $\beta=\beta_s$, the localization of the multi-peak soliton vanishes completely, and the period of the multi-peak soliton become a constant value. Namely,
in this case the multi-peak soliton is converted to a periodic wave [see Fig. \ref{fig.2}(d)]. The corresponding exact expression of the periodic wave is given
\begin{eqnarray}
\label{equ:4}
    E=E_1 \left(\frac{2a^2-2b_1^2}
    {a^2+b_1^2 \cos\sigma_1+b_1 s_1\sin\sigma_1}-1\right),
\end{eqnarray}
where $\sigma_1=2s_1(t+vz)$, $v=2[a^2+2(b_1^2-3a_1 \omega)]\beta_s+2\alpha(\omega+3a_1)$, $s_1=\sqrt{a^2-b_1^2}$.
The periodic wave is formed by the ratio of trigonometric function and exponential function with the period along $t$ direction
$D_t=\pi/\sqrt{a^2-b_1^2}$. It should be pointed out that the periodic wave can exhibit different structures as the period $D_t$ changes.
As $D_t$ increases, i.e., $b_1\rightarrow a$, one can obtain the W-shaped wave trains which is similar to the results in \cite{b6}.

If instead $a^2<b_1^2$, we will observe the impact of higher-order effects on the periodicity of the multi-peak soliton as $\beta\rightarrow\beta_s$.
As shown in Fig. 3, the periodicity of the multi-peak soliton gradually attenuates and the corresponding peak numbers decreases [see Figs. \ref{fig.3}(a)-(c)].
Eventually the periodicity vanishes completely when $\beta=\beta_s$. In this case the multi-peak soliton is converted to a single-peak soliton on
a plane-wave background, i.e., the anti-dark soliton \cite{A} [see Fig. \ref{fig.3}(d)]. The corresponding exact expression of the soliton reads
\begin{eqnarray}
\label{equ:5}
    E=E_1 \left(\frac{2 a^2-2b_1^2}
    {a^2+b_1^2 \cosh\tau_1-b_1 s_2\sinh\tau_1}-1\right),
\end{eqnarray}
where $\tau_1=2s_2(t+vz)$, $s_2=\sqrt{b_1^2-a^2}$. This anti-dark soliton lies on a plane-wave background with single peak
$|E|_{max}^2=(2b_1-a)^2$, Moreover, as $a\rightarrow 0$, this wave will become a standard bright soliton.

\begin{figure*}[htb]
\centering
\includegraphics[height=80mm,width=130mm]{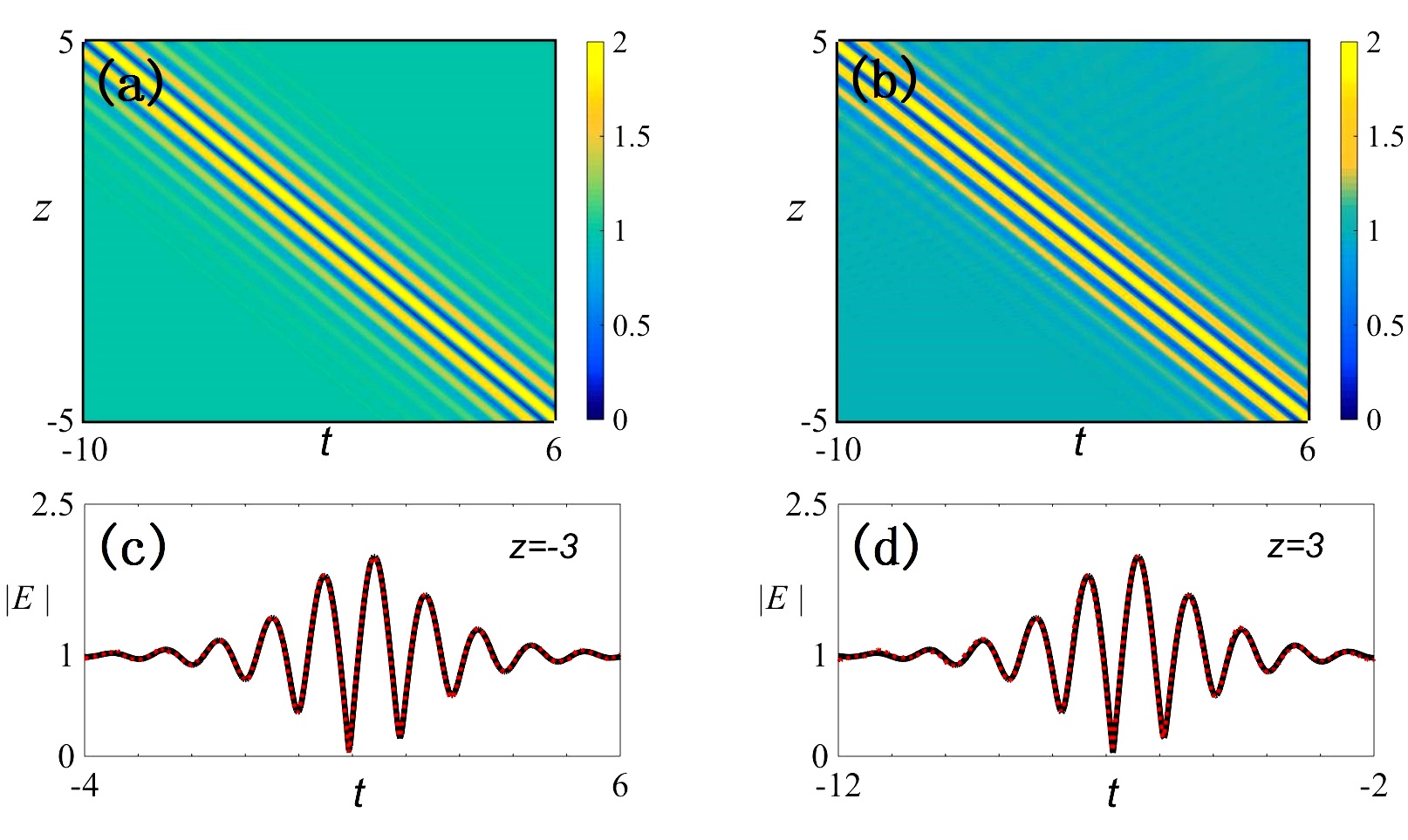}
\caption{(color online) Stability confirmation for the multi-peak solitons $|E(z,t)|$,  (a) the analytical solution, (b) the numerical simulations,
(c) and (d) the corresponding amplitude profiles of the analytical solution (solid black line) and the numerical simulation (dotted red line) at $z=-3,3$.
The existence condition of multi-peak solitons is chosen from Eq. (\ref{equ:c2}) with the parameters: $\beta=0.5\beta_s$, $\alpha=0.5$, $a=1$, $a_1=1$, $b_1=0.5$, $\omega=1$, $k=0.1$.}\label{fig4}
\end{figure*}

\section{Stability of multi-peak solitons}

It is well-known that the stability plays an important role in nonlinear wave realization and application.
One should keep in mind that the multi-peak solitons reported above are on a plane-wave background. The latter, in general, displays the feature of MI,
namely, a small perturbation may distort the wave profiles formed on top of it. In our previous works \cite{l,x2}, we have numerically confirmed the stability of multi-peak solitons and W-shaped solitons (which can be regarded as a limit case of the multi-peak soliton) in the decoupled case of the H-MB system (the Hirota model).

Here we shall study numerically the stability of multi-peak solitons in the H-MB system by the split-step Fourier
method with the initial condition given by the exact solution (\ref{equ:3}) at $z=-5$. The existence condition is chosen from Eq. (\ref{equ:c2})
with the parameters: $\beta=0.5\beta_s$, $\alpha=0.5$, $a=1$, $a_1=1$, $b_1=0.5$, $\omega=1$, $k=0.1$.

Figure \ref{fig4} displays the stability of multi-peak solitons by
comparison the analytical solution (\ref{equ:3}) $E$ with the corresponding numerical simulation.
Interestingly, the numerical result [Fig. \ref{fig4} (b)] shows clearly that the multi-peak soliton can propagate in a stable manner against
the MI and ineradicable numerical deviations.
In particular, we compare the numerical result with the analytical solution by the
corresponding amplitude profiles at different $z$. It shows in Figs. \ref{fig4} (c) and (d) that the numerical result (dotted red line) is in good agreement with the
analytical solution (solid black line) at $z=-3,3$. These results confirm numerically
the stability of multi-peak solitons in the general H-MB system, which could be useful for the corresponding experimental observation in an erbium-doped fiber system.

\section{multi-peak solitons in degenerate cases}
It is evident that the H-MB system contains three special cases:
\rmnum{1}) the NLS-MB model with $\beta=0$ (thus without higher-order effects);
\rmnum{2}) the MB model with $\alpha=0$, $\beta=0$;
\rmnum{3}) the decoupled case (i.e., the Hiorta model \cite{Hirota}) with $k=0$.

Remarkably, our multiparametric solution (\ref{equ:3}) with a general form contains the above three special degenerate cases.
Thus the similarity is that the multi-peak solitons can exist in the three different nonlinear contexts.
Since the H-MB system can be regarded as a coupled system of the Hirota and MB models, we will in this section
compare and discuss the similarity and difference of the multi-peak solitons in the coupled case and two special degenerate cases (the Hirota and MB cases).

Additionally, we omit the results of the NLS-MB model since the property of the multi-peak solitons has been revealed in \cite{b6}.
Nevertheless, one should note that the multi-peak soliton in the NLS-MB system is induced by the coupled MB components, which cannot exist in the standard NLS system.

Here we first study the multi-peak solitons in the MB system (case \rmnum{2}). It is obvious that the multi-peak mode can be extracted from Eq. (\ref{equ:3}) with $\alpha=0$, $\beta=0$, $k=0$ (thus $v_2=0$).
We remark that only a peculiar kind of multi-peak solitons, i.e., the stationary multi-peak soliton ($v_2=0$, $v_1=0$) can exist in the MB system, which is different from the one in the coupled H-MB model.

We then study multi-peak solitons in the scalar Hirota model in case \rmnum{3}. The multi-peak mode is extracted directly from Eq. (\ref{equ:3}) with the parameter condition:
\begin{eqnarray}
\alpha+\beta q-4a_1 \beta=0.
\end{eqnarray}
It shows, in the Hirota case, the impact of higher-order effects on the multi-peak solitons is similar to the results in the H-MB system. Namely, the
higher-order effects impact the velocity when $\beta$ and $q$ are two independent parameters (f.i., $\alpha=4a_1 \beta-q\beta$);
while higher-order effects impact the periodicity and localization when $q$ is expressed as a function of $\beta$, i.e., $q=4a_1-\alpha/\beta$. Specifically,
if $\beta=\alpha/(6a_1)$, the multi-peak solitons can be converted to the periodic wave ($a^2>b_1^2$), and the anti-dark soliton ($a^2<b_1^2$).
One should also note that in the special condition, i.e., $q=0$, the multi-peak soliton in the Hirota model has been obtained in \cite{a2}.

Thus, what is the difference between the multi-peak solitons in the H-MB and Hirota models? We note that the amplitude parameter $k$ in the H-MB system can suppress
the impact of higher-order effects on the periodicity and localization of multi-peak solitons.
Specifically, if $k=2\alpha n$, implying $q=-4a_1$ (thus the $q$ is unrelated to $\beta$), the constraint condition (\ref{equ:c2}) reduce to the constraint condition (\ref{equ:c1}). In this case, the higher-order effects only impact the velocity of multi-peak solitons in the H-MB systm. In contrast, the higher-order effects impact both the velocity and transitions of multi-peak solitons in the Hirota model, and the suppression of the impact of higher-order effects on the periodicity and localization of multi-peak solitons cannot be obtained.

\section{Conclusion}
In summary, we have studied the characteristics of multi-peak solitons on a plane-wave background in an erbium-doped fiber system with higher-order effects governed by
the H-MB model. The important properties of multi-peak solitons induced by the higher-order effects, such as the velocity changes,
localization or periodicity attenuation, and state transitions, were analyzed in detail. The stability of the multi-peak solitons was verified numerically by
means of direct split-step Fourier method.
Moreover, we compared and discussed the similarity and difference of multi-peak solitons in special degenerate cases of the H-MB system with a general existence condition.
These results will enrich the understanding of the characteristics of multi-peak solitons induced by the higher-order effects in the
H-MB and other nonlinear contexts governed by higher-order NLS models.

\section*{ACKNOWLEDGEMENTS}
We appreciate the anonymous referees for their constructive comments to improve the paper. C. Liu also thanks Li-Chen Zhao for his helpful discussion.
This work has been supported by the National Natural Science Foundation of China (NSFC)(Grant Nos. 11475135, 11547302), and NWU graduate student innovation funds (Grant No. YZZ15085).


\begin{thebibliography}{99}
\bibitem{n} A. I. Maimistov and A. M. Basharov, Nonlinear Optical Waves (Springer-Verlag, Berlin, 1999);
\bibitem{n1} S. Kakei and J. Satsuma, J. Phys. Soc. Jpn. \textbf{63} (1994) 885.
\bibitem{n2} K. Porsezian and K. Nakkeeran, J. Mod. Opt. \textbf{42} (1995) 1953;
K. Porsezian J. Mod. Opt. \textbf{47} (2000) 1635; K. Porsezian, P. Seenuvasakumaran, and R. Ganapathy, Phys. Lett. A \textbf{348} (2006) 233.
\bibitem{n3} M. S. M. Rajan, A. Mahalingam, A. Uthayakumar, J. Opt. \textbf{14}, 105204 (2012).
\bibitem{h1} K. Porsezian and K. Nakkeeran, Phys. Rev. Lett. 74 (1995) 2941; K. Nakkeeran and K. Porsezian, J. Phys. A \textbf{28}, 3817 (1995); K. Nakkeeran and K. Porsezian, Opt. commun. \textbf{123}, 169 (1996).
\bibitem{e1} M. Nakazawa, E. Yamada, and H. Kubota, Phys. Rev. Lett. \textbf{66}, 2625 (1991).
\bibitem{e2} M. Nakazawa, E. Yamada, and H. Kubota, Phys. Rev. A \textbf{44}, 5973 (1991).
\bibitem{e3} M. Nakazawa, Y. Kimura, K. Kurokawa, and K. Suzuki, Phys. Rev. A \textbf{45}, R23 (1992).
\bibitem{e4} M. Nakazawa, K. Suzuki, Y. Kimura, and H. Kubota, Phys. Rev. A \textbf{45}, R2682 (1992).
\bibitem{r1} N. Akhmediev, J. M. Soto-Crespo, and A. Ankiewicz, Phys. Lett. A \textbf{373}, 2137 (2009 ); N. Akhmediev, A. Ankiewicz, and M. Taki,\textit{ibid}.
             \textbf{373}, 675 (2009 ); N. Akhmediev, J. M. Soto-Crespo, and A. Ankiewicz, Phys. Rev. A \textbf{80}, 043818 (2009).
\bibitem{r2} J. M. Dudley, F. Dias, M. Erkintalo, and G. Genty, Nat. Photon. \textbf{8}, 755 (2014); N. Akhmediev, J. M. Dudley, D. R. Solli, and S. K. Turitsyn, J. Opt, {\bf15}, 060201 (2013); N. Akhmediev, B. Kibler, F. Baronio, et al, J. Opt, {\bf18}, 063001 (2016).
\bibitem{b1} J. S. He, S. W. Xu, and K. Porsezian, J. Phys. Soc. Jpn. \textbf{81} (2012) 033002.
\bibitem{b2} J. S. He, S. W. Xu, and K. Porsezian, Phys. Rev. E \textbf{86}, 066603 (2012).
\bibitem{b3} C. Z. Li, J. S. He, and K. Porsezian, Phys. Rev. E \textbf{87}, 012913 (2013).
\bibitem{b4} R. Guo and H. Q. Hao, Commun. Nonlinear Sci.Numer. Simulat. \textbf{19}, 3529 (2014); Ann. Phys. \textbf{344}, 10 (2014).
\bibitem{b5} L. Wang, X. Li, F. H. Qi, and L. L. Zhang, Ann. Phys. \textbf{359}, 97 (2015).
\bibitem{b6} Y. Ren, Z. Y. Yang, C. Liu, and W. L. Yang, Phys. Lett. A \textbf{379}, 2991 (2015);
             L. Duan, Z. Y. Yang, C. Liu, and W. L. Yang, arXiv:1601.03145 (2016).
\bibitem{b7} L. Wang, Y. J. Zhu, J. H. Zhang, T. Xu, F. H. Qi, and Y. S. Xue, J. Phys. Soc. Jpn. \textbf{85}, 024001 (2016).
\bibitem{t1} G. P. Agrawal, J. Opt. Soc. Am. B \textbf{28}, A1 (2011).
\bibitem{t2} G. P. Agrawal, \emph{Nonlinear Fiber Optics} (4th Edition, Acdemic Press, Boston, 2007).
\bibitem{t3} N. Akhmediev and A. Ankiewicz, \emph{Solitons: Nolinear Pulses and Beams} (Chapman and Hall, London, 1997).
\bibitem{t4} J. Yang, \emph{Nonlinear Waves in Integrable and Nonintegrable Systems} (SIAM, Philadelphia, 2010).
\bibitem{s1} A. Ankiewicz and N. Akhmediev, Phys. Lett. A {\bf378}, 358 (2014).
\bibitem{s2} A. Chowdury, D. J. Kedziora, A. Ankiewicz, and N. Akhmediev, Phys. Rev. E \textbf{90}, 032922 (2014).
\bibitem{m} R. Guo, B. Tian, X. L\"{u}, H. Q. Zhang, and W. J. Liu Comput. Math. Phys. \textbf{52} (2012) 565;
R. Guo, H. Q. Hao and L. L. Zhang, Mod. Phys. Lett. B \textbf{27} (2013) 1350130.
\bibitem{a1} A. Chowdury, D. J. Kedziora, A. Ankiewicz, and N. Akhmediev, Phys. Rev. E \textbf{91}, 032928 (2015).
\bibitem{a2} A. Chowdury, A. Ankiewicz, and N. Akhmediev, Proc. R. Soc. A \textbf{471}, 20150130 (2015).
\bibitem{w} L. Wang, J. H. Zhang, Z. Q. Wang, C. Liu, M. Li, F. H. Qi, and R. Guo, Phys. Rev. E \textbf{93}, 012214 (2016).
\bibitem{w1} L. Wang, J. H. Zhang, C. Liu, M. Li, and F. H. Qi, arXiv:1603.01456.
\bibitem{l} C. Liu, Z. Y. Yang, L. C. Zhao, L. Duan, G. Y. Yang, and W. L. Yang, arXiv:1603.04554.
\bibitem{x1} A. Ankiewicz, J. M. Soto-Crespo, and N. Akhmediev, Phys. Rev. E {\bf81}, 046602 (2010);
             A. Ankiewicz, J. M. Soto-Crespo, M. A. Chowdhury, and N. Akhmediev, J. Opt. Soc. Am. B {\bf30}, 87 (2013);
             G. Y. Yang, L. Li, and S. T. Jia, Phys. Rev. E {\bf85}, 046608 (2012);
             Y. S. Tao and J. S. He, Phys. Rev. E {\bf85}, 026601 (2012).
\bibitem{x2} C. Liu, Z. Y. Yang, L. C. Zhao, and W. L. Yang, Phys. Rev. E \textbf{91} (2015) 022904.
\bibitem{x3} A. Ankiewicz, Y. Wang, S. Wabnitz, and N. Akhmediev, Phys. Rev. E {\bf89}, 012907 (2014).
\bibitem{x4} A. Chowdury, D. J. Kedziora, A. Ankiewicz, and N. Akhmediev, Phys. Rev. E \textbf{91}, 022919 (2015).
\bibitem{x5} L. C. Zhao, C. Liu, Z. Y. Yang, Commun. Nonlinear Sci. Numer. Simulat. {\bf20}, 9 (2015).
\bibitem{A} Yu. S. Kivshar, Phys. Rev. A \textbf{43} (1991) 1677; Yu. S. Kivshar and V. V. Afanasjev, Phys. Rev. A \textbf{44} (1991) 1446(R).
\bibitem{Hirota} R. Hirota, J. Math. Phys. \textbf{14}, 805 (1973).



\end{thebibliography}
\end{document}